\documentclass[aps,prl,preprint,groupedaddress,showpacs]{revtex4}

\usepackage{graphicx}
\usepackage{dcolumn}
\usepackage{bm}

\begin{document}


\title{Constraints on Trilinear Coupling Constant $A_{0}$ 
in minimal Supergravity Model}


\author{Jun Tabei}
 \email{jun@hep.phys.waseda.ac.jp}
\affiliation{%
Department of Physics, Waseda University, Tokyo 169, Japan
}%

\author{Hiroshi Hotta}
 \email{hotta@hep.phys.waseda.ac.jp}
\affiliation{
Institute of Material Science and Technology, Waseda University, 
Tokyo 169, Japan
}%


\date{\today}

\begin{abstract}
Constraints on the common trilinear coupling constant $A_{0}$ 
are obtained in the minimal supergravity model. 
$A_{0}$ is usually given at GUT scale $M_{X}$ by-hand, 
and has been regarded as one of the five arbitrary parameters 
in this model. 
However, $B(M_{Z})$ as the common bilinear coupling constant 
$B$ at the electroweak scale $M_{Z} \simeq 91$(GeV) is fixed by the 
shape of the Higgs potential. 
With the renormalization group equations, 
this $B(M_{Z})$ is evolved to $B_{0}$ at $M_{X}$ where 
a well-known relation exists between 
the evolved $B_{0}$ and $A_{0}$ as $B_{0} = (A_{0}-1) m_{0}$. 
Therefore, $A_{0}$ is fixed as a function of 
the other arbitrary parameters. 
In other words, the number of arbitrary parameters 
are reduced by this method. 
Additionally, large $\tan{\beta} > 40$ is forbidden 
by the conditions at $M_Z$. 
\end{abstract}

\pacs{11.10.Hi, 12.60.Jv, 04.70.-S, 12.10.-g}

\maketitle



It has been considered that the minimal supergravity model 
contains five arbitrary parameters 
($A_{0}, m_{0}, m_{1/2}, \tan{\beta}, \mbox{sig}(\mu)$), 
where 
$A_{0}$ stands for the common trilinear coupling constant of 
the scalar fields at the breakdown scale 
$M_{X} \simeq 2 \times 10^{16}$(GeV) of the supersymmetry. 
$\tan{\beta}$ is defined as the ratio $\tan{\beta}=v_{u}/v_{d}$ 
between the vacuum expectation 
values(VEVs) of the two Higgs doublets of the u-quark$(v_{u})$ 
and of the d-quark$(v_{d})$ sectors as usual. 
$m_{0}$ and $m_{1/2}$ are the common scalar and the common 
gaugino mass at $M_{X}$, respectively. 
$\mbox{sig}(\mu)$ means the sign of the Higgs mixing parameter $\mu$. 
There exists a well-known relation between $A_{0}$ and 
the bilinear coupling constant $B_{0}$ at $M_{X}$ 
in this model\cite{Nilles}, 
\begin{equation}
B_{0}=(A_{0}-1)m_{0} \ \ , 
\label{eq:b0}
\end{equation}
however, of this relation have never been thought as a 
constraint on $A_{0}$, because both $B_{0}$ and $m_{0}$ have been 
also regarded as arbitrary parameters at $M_{X}$. 
Nevertheless, $B(M_{Z})$ as the bilinear coupling constant $B$ 
at the electroweak scale $M_{Z} \simeq 91$(GeV) 
is fixed by the shape of the Higgs potential. 
Once the bilinear coupling constant $B$ at $M_{Z}$ is given, 
corresponding $B_{0}$ (at $M_{X}$) is obtained by the evolution 
of the renormalization group equations(RGEs) of this model. 
Therefore, the conditions on $B(M_{Z})$ fix $A_{0}$ at $M_{X}$ 
with Eq.(\ref{eq:b0}). 
Furthermore, $\tan{\beta}$ is restricted 
by the conditions at $M_{Z}$ to realize the appropriate structure 
of the vacuum at $M_{Z}$. 
Derived constraints on $A_{0}$ and $\tan{\beta}$ 
are presented in the following.

The boundary conditions of the RGEs to analyse the behaviors 
of the parameters are given here. All the RGEs are 
at one-loop level\cite{Castano},\cite{Arason}. 

\begin{itemize}

\item[(1)]
{\bf Conditions at $M_{X}$}
\begin{equation}
B_{0}=(A_{0}-1)m_{0} 
\ \ . \label{eq:a0} 
\end{equation}
Note that the breakdown scale $M_{X}$ of the supersymmetry is made 
equal to the energy-scale of the gauge grand unification 
in this paper. 

\item[(2)]
{\bf Conditions at $M_{Z}$}
\\
The boundary conditions of 
the gauge coupling constants are given at $M_{Z}$\cite{Particle}. 
On the Higgs sectors at $M_{Z}$, 
\begin{equation}
\mu^2=-\frac{m_Z^2}{2}+
\frac{m_{h_d}^2-m_{h_u}^2\tan^2{\beta}}{\tan^2{\beta} -1} 
\ \ , 
\end{equation}
\begin{equation}
B(M_{Z})=-\frac{1}{2\mu}\left(
m_{h_d}^2-m_{h_u}^2+2\mu^2\right)
\sin{2\beta}
\ \ , \label{eq:bmz}
\end{equation}
where, $m_{h_d}$ and $m_{h_u}$ stand for the soft-breaking mass 
parameters of the d-type and the u-type quark sectors 
in each Higgs doublet at $M_{Z}$, respectively. 
Since the sign of $\mu$ can be positive or negative, both of these 
two situations will be taken place. 
The vacuum expectation values(VEVs) of the Higgs potential 
at $M_{Z}$ should keep the appropriate positivity. 
Moreover, the squared CP-odd Higgs mass also should be positive. 
Furthermore, 
the squared soft-breaking masses of the squarks and sleptons 
have to be positive, 
in other words, the squarks and sleptons should be kept in no VEV. 

\item[(3)]
{\bf Other conditions}
\\
The decoupling energy-scale $M_{SUSY}$ of the supersymmetric 
region is fixed to $1$(TeV) as usual\cite{Amaldi}. 
On larger three(top, bottom, and tau) Yukawa 
coupling constants, their values are referred from the 
latest issue of the Particle Data Group\cite{Particle}, 
and fixed on their self mass-shell scales, respectively. 
The other(light) quarks and leptons are neglected. 
\end{itemize}

The behaviors of $A_{0}$ are numerically derived by 
the RGE analysis under these conditions.

Figures 1a and 1b show $A_{0}$(solid curves) 
at $m_{1/2}/m_{0}$=0.5, 1, and 2. 
The authors found $A_{0}$ does not depend 
on each absolute values of $m_{0}$ or $m_{1/2}$, 
but only depends on the dimensionless ratio of $m_{1/2}/m_{0}$, 
if both of $m_{0}$ and $m_{1/2}$ are larger than 
the electroweak scale $M_{Z}$. 
The rightsides of the dashed curves plotted 
for $0.1 \leq m_{1/2}/m_{0} \leq 10$ 
are excluded by the conditions at $M_{Z}$. 
Allowed region of $A_{0}$ with $\mu > 0$ 
is very narrow as shown in Fig. 1a. 
Especially, $A_{0}$ is fixed to $1.5$ independent of 
the ratio $m_{1/2}/m_{0}$ when $\tan{\beta} \simeq 7$. 
With $\mu < 0$, $A_{0}$ is always larger than 2, 
and restricted in shallowly U-shaped band 
on the support of $2 \leq \tan{\beta} \leq 40$ 
as shown in Fig. 1b. 
The bottom of $A_{0}$ is always around $\tan{\beta}=15$ 
with $\mu < 0$. 
From the qualitative point of view, the behaviors of $A_0$ 
at $M_{X}$ directly reflect ones of $B(M_{Z})$ at $M_{Z}$ 
through the RGEs. 

As the results of the RGE analysis, 
$A_{0}$ is fixed as a function of the other parameters. 
Such tight constraints on $A_{0}$ are 
derived from the common bilinear coupling constant $B(M_{Z})$ 
in the Higgs sector at $M_{Z}$ and the relation 
$B_{0}=(A_{0}-1)m_{0}$ at $M_{X}$ scale. 
Additionally,  $\tan{\beta}$ is restricted as smaller 
than about 40 almost independent of the other free parameters. 
This restriction on $\tan{\beta}$ stems from the conditions 
at $M_{Z}$, where the squared soft-breaking masses of the 
squarks and sleptons (especially, of the stau) should be 
positive, and the Higgs potential must keep the appropriate 
VEVs. 




\begin{acknowledgments}
The authors thank H. Oka for a lot of useful advice. 
\end{acknowledgments}

%



\pagebreak


%


\begin{figure}
\centering
\includegraphics[width=12cm]{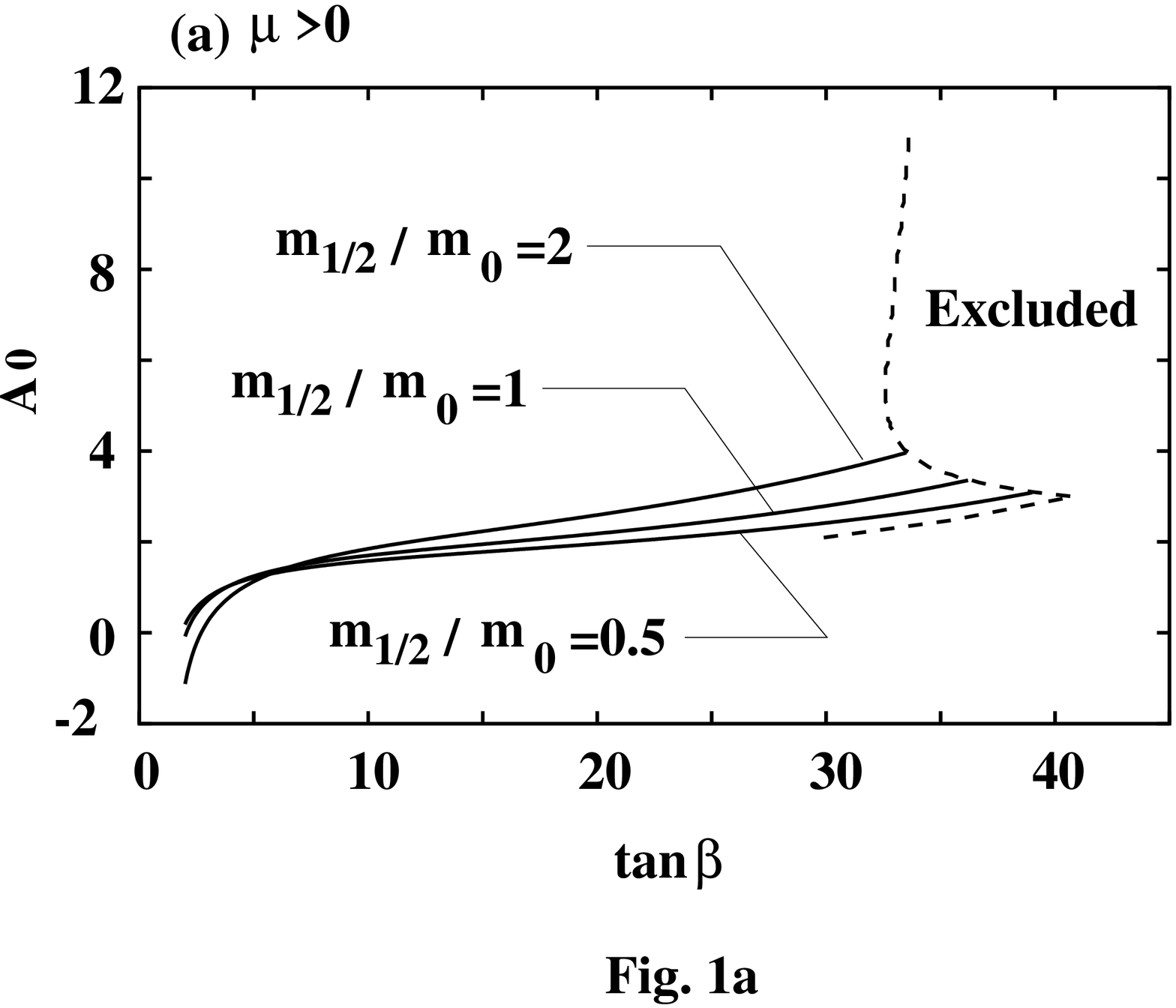}
\ \\
\ \\
\includegraphics[width=12cm]{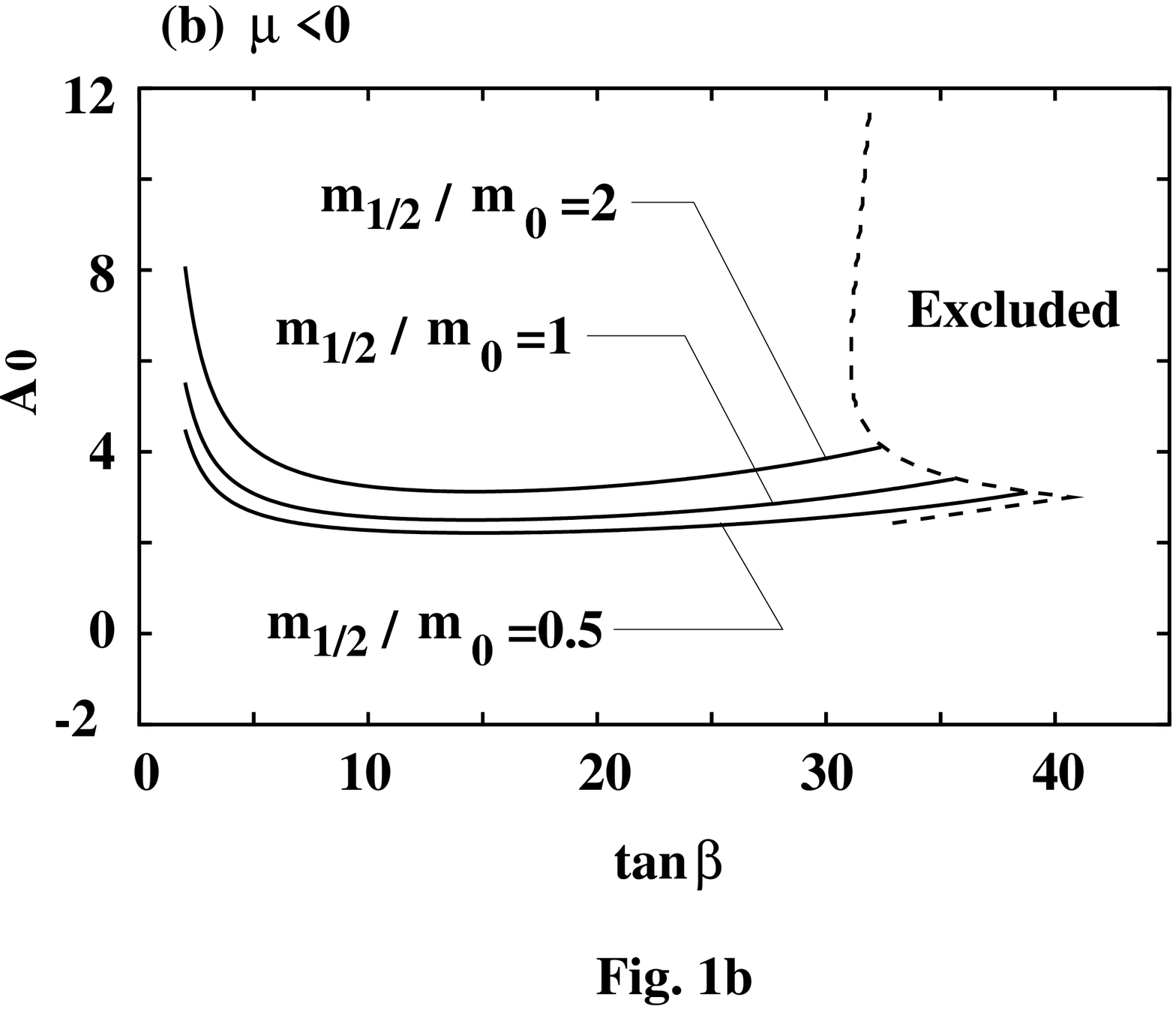}
\caption{\label{fig:1} 
$A_{0}$ as functions of $\tan{\beta}$ 
at $m_{1/2}/m_{0} = 0.5$, 1, and 2. 
with (a) $\mu > 0$ and (b) $\mu < 0$. 
The rightsides of the dashed curves are excluded. 
}
\end{figure}


%

\end{document}